\documentclass[pre,aps,groupedaddress,amsmath,twocolumn]{revtex4}
\usepackage[dvips]{graphicx}
\usepackage{dcolumn}
\usepackage{bm}
\usepackage{epsfig}

\begin{document}

\title{Charge and spin correlations of a one dimensional electron gas on the
  continuum}

\author{Michele Casula,$^{1,2}$ and Gaetano Senatore$^{3}$}
\affiliation{
$^1$ International School for Advanced Studies (SISSA) Via Beirut 2,4
  34014 Trieste, Italy \\
$^2$ INFM Democritos National Simulation Center, Trieste, Italy \\
$^3$ Dipartimento di Fisica Teorica dell' Universit\`a di Trieste, 
Strada Costiera 11, 34014 Trieste, Italy
} 

\date{}

\begin{abstract}
  We present a variational Monte Carlo study of a model one dimensional
  electron gas on the continuum, with long-range interaction ($1/r$ decay).
  At low density the reduced dimensionality brings about pseudonodes of the
  many-body wavefunction, yielding non-ergodic behavior of naive Monte Carlo
  sampling, which affects the evaluation of pair correlations and the related
  structure factors.  The problem is however easily solved and we are able to
  carefully analyze the structure factors obtained from an optimal trial
  function, finding good agreement with the exact predictions for a
  Luttinger-like hamiltonian with an interaction similar to the one used in
  the present study.
\end{abstract}

\maketitle

\section{introduction}

Electrons in low dimensions, as found in modern semiconductor
devices\cite{nano89}, are greatly affected by correlation effects that may
dramatically change their behavior and bring about new phenomena and phases.
Technological advances have reached the point where it is now possible to
fabricate high mobility systems in which the electrons move essentially free
and are intentionally confined at the nanoscale in suitable geometries.
Though the details of the devices still play a role at the quantitative
level\cite{botti}, to a good approximation some of these systems provide a
very close realization of simple electron gas models. We should emphasize that
this is true only in low dimension and the experimental realization of a clean
3-dimensional (3D) electron gas (EG) in the quantum regime is in fact still
out of sight, as is the long sought 3D Wigner crystal.  Here, we focus our
attention on a simple 1D model with long-range interaction, i.e.,
a quantum wire, with the aim of examining its structural properties.

A large variety of models have been proposed over the years for quantum wires.
One may roughly distinguish two classes of systems, depending whether the wire
is obtained from the confinement of a 3D\cite{harm,hard} or of a
2D\cite{2dharm,2dhard,2dcoul} electron gas; or one may focus on the nature of
the confinement, which can be harmonic\cite{harm,2dharm}, hard-wall
like\cite{hard,2dhard} or Coulombic\cite{2dcoul}. In all these cases, the
effective electron-electron interaction in the lowest subband of the
transverse motion turns out to have a $1/r$ long-range tail, whereas the
details of the confinement only affect the behavior at short range, which
displays anyhow a divergence milder than $1/r$, if any. Thus, if the interest is
in the effect of the long range tail of the interaction, one may well resort to
ad-hoc choices of the interaction\cite{schulz} or even to tight-binding models
yielding Hubbard-like Hamiltonians with long range intersite
interaction\cite{parola,capponi}; however, this last choice may
introduce\cite{capponi} processes (such as the Umklapp scattering) not present
in continuum models. One should also recall that, for interacting Fermions in
1D, the familiar Fermi-liquid paradigm must be abandoned in favor of the
Tomonaga-Luttiger liquid concept\cite{lutt}.

The above model wires have been studied at $T=0$ employing several techniques
ranging from DFT treatments\cite{2dcoul} to dielectric-formalism approaches of
the STLS type\cite{harm,bloch}, from exact diagonalization\cite{exact} to
density matrix renormalization group techniques\cite{parola}, to bosonization
techniques\cite{schulz}--the last approach being able to provide exact results
for some properties if a linearized kinetic energy dispersion is assumed.
Investigations of the homogeneous 1DEG on the continuum, performed with either
the variational (VMC) or the diffusion Monte Carlo (DMC), are also
available\cite{malathesis,malatesta,casulat}. Contrary to STLS
predictions\cite{bloch} for the same model, DMC
simulations\cite{malathesis,malatesta} yield no Bloch instability (spontaneous
magnetization of the system), in accordance with the Lieb and Mattis
theorem\cite{lieb}. They predict pair correlations that are consistent with a
quasi Wigner crystal (strong charge correlations resulting in a divergent peak
in the charge structure factor at $4k_F$\cite{schulz}). However, they also
appear to imply the presence of quasi antiferromagnetic order, i.e., a
divergence of the spin structure factor at $2 k_F$ in the unpolarized state, a
result at variance with the findings of the bosonization
technique\cite{schulz}. Rather than to the differences between
models\cite{malatesta}, such a discrepancy is likely to be due to the
ergodicity problems encountered in both DMC and VMC simulations at strong
coupling\cite{malathesis,casulat}, as exchange between opposite spin electron
is frozen out by the strong Coulomb repulsion.  While in the DMC method a
solution of such a problem is not straightforward\cite{casulat}, in VMC
simulations one may device fairly simple ways of restoring
ergodicity\cite{malathesis,casulat}, thus establishing the true nature of spin
correlations, as we show below.

\section{Model and wavefunction}
The Hamiltonian of our model quantum wire\cite{harm} is, 
in units in which $a_0^*=\hbar^2\epsilon/(m^* e^2)=1$ and $e^2/(2 \epsilon
a_0^*)=1$, 
\begin{eqnarray}
\label{ham}
H = -\sum_{i=1}^N \nabla^2_i + 2 \sum_{i<j} V_b(|x_i-x_j|),\\
V_b(x)= (\sqrt{\pi}/{2 b}) {\rm exp} \left( {x^2}/{4 b^2} \right) {\rm
  erfc} \left({ |x|}/{2 b} \right),  
\label{pot}
\end{eqnarray}
with $b$ a measure of the wire width. Above, $\epsilon$ is the
dielectric constant of the semiconductor and $m^*$ the effective mass of the
carriers.  The pair interaction $V_b(x)$, which is finite at the origin,
$V_b(0)=\sqrt{\pi}/(2b)$ and decays as $1/x$ for $x\gg b$, is obtained from an
harmonic confinement of the 3DEG, after projection on the lowest subband of
transverse motion\cite{harm}; it thus provides a good approximation to the 3D
system at low density $\rho$, i.e., $r_s =1/2\rho \gg \pi b/4$, with the
density parameter $r_s$ also providing an estimate of the Coulomb coupling, as
ratio between average potential and kinetic energies.

We use VMC simulations and focus on the ground state properties of a thin wire
($b=0.1$) on a fairly wide coupling range ($1\leq r_s\leq 10$), thus
considering only unpolarized states\cite{lieb,malatesta}. To this end, we
resort to a Slater-Jastrow wavefunction\cite{cep78} $\Psi_T=J D^\uparrow
D^\downarrow$, with $D^\sigma$ a determinant of $N^{\sigma}$ plane waves and
$J=\exp [-\sum_{i<j} u(|x_i-x_j|)]$, with $u(x)$ a two-body pseudopotential to
be optimized. This is the simplest correlated wavefunction for an unpolarized
Fermion state with homogeneous density and can be further improved with the
inclusion in the Jastrow factor $J$ of higher order
pseudopotentials\cite{emc}.  Our goal here is a careful finite-size scaling
extrapolation of structural properties, to obtain the long-range behavior of
pair correlation functions.  Thus, to reduce the finite size bias we keep
$N^\uparrow$= $N^\downarrow$ odd, in order to avoid shell degeneracy effects,
use periodic boundary conditions and Ewald-sum\cite{cep78,malathesis,casulat}
the pair potential $V_b(x)$.

We remind the reader that in 1D the nodes of the Fermionic {\it ground state}
(of given symmetry) are know exactly\cite{ceperley1d}, as they are fully
determined by exchange (anti)symmetry, and coincide in particular with those
of the above wavefunction for the unpolarized state.  Thus, DMC provides in
principle exact energies\cite{malathesis,malatesta,casulat} that may serve as
benchmark for the VMC simulations when optimizing the Jastrow factor $J$.  As
starting point for our simulations we take a two-body pseudopotential of the
RPA type \cite{jastrowrpa,cep78}, which in Fourier space reads
\begin{equation}
{2\rho} \tilde{u}_{RPA}(k)=  
  -S_0(k)^{-1}+\sqrt{S_0(k)^{-2}+ 
{4 \rho \tilde{V}_b(k)}/{k^2}} , 
\label{u_rpa}
\end{equation}
with $S_0(k)=(k/2k_F)\theta(2k_F-k)+\theta(k-2k_F)$ the structure factor of a
non interacting 1DEG,  
\begin{equation}
\tilde{V}_b(G)=E_1(b^2 G^2) \exp(b^2 G^2)
\end{equation}
the Fourier transform of the interaction, and $\theta(x)$ and $E_1(x)$
respectively the  Heaviside and the exponential  integral functions.
\begin{figure}[t]
\null\vspace{-3mm}
\epsfxsize=85mm
\centerline{
\epsffile{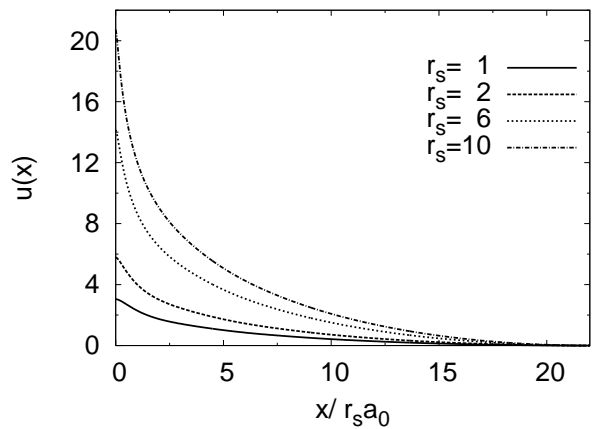}}
\caption{RPA  pseudopotential for $b=0.1$ and  $r_s=1$ (solid line), 
  $r_s=2$ (dashed line), $r_s=6$ (dotted line), $r_s=10$ (dot-dashed line).}
\label{rpa}
\end{figure}

The repulsive nature of the pair interaction $V_b(x)$ is directly reflected in
the pseudopotential, which is shown in Fig.~\ref{rpa}. The pseudopotential is
repulsive and the repulsion increases appreciably with decreasing the
density (increasing the $r_s$).  Thus, as the density is lowered electrons are
kept apart more and more effectively and this results in a quasi long range
order which can be described as a quasi Wigner Crystal\cite{schulz}.  Though
the pseudopotential remains finite at contact, the Jastrow factor becomes
exponentially small yielding what may be seen as pseudonodes of the
wavefunction. These pseudonodes have no particular effect on like spin
electrons, as the wavefunction is already vanishing at contact for such
electrons and most importantly particle exchanges are explicitly summed over
in the determinants. On the contrary, the effect on opposite spin electrons
when combined with the reduced dimensionality may become dramatic.  In a
random walk in configuration space with importance sampling given by the
Slater-Jastrow wavefunction, the RPA pseudopotential tends to freeze out the
exchange between opposite spin electrons and may cause ergodicity problems,
which show up in spin correlations\cite{malathesis,casulat}.  Evidently, it is
not only the presence of the pseudonodes to cause problems but also the {\it
  slope} with which the pseudonodes are approached, on the scale of the
interparticle distance. When such a slope becomes sufficiently large, naive
algorithm may become inefficient in sampling inequivalent pockets in
configuration space, delimited by the pseudonodes.

\section{Ergodicity and VMC optimization}

Ergodicity becomes an issue in the present context because in 1D there is no
exchange without crossing, for opposite spin electrons.  With increasing the
coupling, the maximum displacement $\delta$ has to be made smaller (on the
$r_s$ scale), to keep the acceptance high, and this reduces the exchange
frequency $\nu=\rm spin \, flips/total\, moves$ substantially.  The problem is
very serious in DMC simulations, as very small displacements are
necessary\cite{ceperley}, to keep the time step error small, and a direct
sampling of spin exchange is not straightforward.  On the contrary, in the VMC
method it is relatively easy to devise effective ways of keeping $\nu$ above a
suitable threshold ($\approx 0.01$) ensuring ergodicity.  The simplest choice
is to work at acceptance appreciably lower than 50\%, i.e., at larger $\delta$
.  Other possible choices involve either umbrella sampling
/reweighting\cite{reweighting}, i.e.  the use of configurations generated from
a guidance function $\Psi_G$ less repulsive than the trial function $\Psi_T$
considered above, or the use of a generalized Monte Carlo move, involving the
direct sampling of exchange between opposite spin electrons\cite{malathesis}.
One may also consider combinations of the above.  In Tab.~\ref{exchtab}, we
focus our attention on the first two choices, for the worst case examined
here, i.e., $r_s=10$.  It is clear that both the large $\delta$ and the
$\Psi_G$ recipes are effective in keeping $\nu$ close to value
$\approx 0.01$ found at small couplings, $r_s \approx 1$, where there are
no apparent signals of nonergodicity. Indeed we find that keeping $\nu\approx
0.01$ the noisy patterns previously found in the spin structure
factor\cite{malathesis,casulat} disappear. On the other hand, the energy does
not seem to depend on $\nu$ and this is likely due to the fact that at strong
coupling different spin configurations are almost degenerate. In the following
we restrict our attention to the large $\delta$ prescription which has the
advantage of maximum simplicity.
\begin{table}[t]
\begin{ruledtabular}
\begin{tabular}{c|c|c|c|c}
           \multicolumn{1}{c}{Sampling}
         & \multicolumn{1}{c}{$\delta/2r_s$}
         &\multicolumn{1}{c}{$\mathcal{A}$}
         &\multicolumn{1}{c}{$\nu$}
         &\multicolumn{1}{c}{$E$}
           \\
\hline
$\Psi_T$   & 1.3  & 0.49  & $6.06 \cdot 10^{-4}$   & -0.474825(9)   \\
\hline                                                                    
$\Psi_T$   & 3.0  & 0.24  & $1.03 \cdot 10^{-2}$   & -0.474827(9)   \\
\hline                                                                    
$\Psi_G$   & 1.3  & 0.51  & $7.36 \cdot 10^{-3}$   & -0.474832(8)   \\    
\end{tabular}
\end{ruledtabular}
\caption{Exchange frequency $\nu$ at $r_s=10$ and $N=22$.   $\delta$,
         $\mathcal{A}$ and $E$ are the maximum displacement of the MC move,
 the average acceptance, and the energy per particle. In all cases an
 optimized (scaled) RPA pseudopotential (see text) is used.}
\label{exchtab}  
\end{table}

To get an accurate description of correlation functions we systematically
optimize $\Psi_T$, employing the variance minimization method\cite{umrigar}.
As the plane-wave determinants provide the exact nodes for our 1D system, we
only need to optimize the Jastrow factor $J$, i.e., the two-body
pseudopotential $u(x)$, for which we have considered both a scaled RPA form
$u_\alpha(x)=\alpha u_{RPA}(x)$ and a systematic expansions in
terms of Chebyshev polynomials
\begin{equation}
u(x) = \sum_{m =1}^{m_{max}} \alpha_m T_{2m}
 ( 2|x|/L-1),
\label{tchebymin}
\end{equation}
satisfying the continuity of the first derivative at the edge of the {\sl
  simulation box}, $u^\prime(L/2)=0$.  The results of the optimization are
summarized  in Tab.~\ref{compenergy}, in terms of the correlation energy
$E_c=E_0-E_{HF}$, with $E_0$ and $E_{HF}$ the exact (DMC) and Hartree-Fock
energies.
\begin{table}[b]
\begin{ruledtabular}
\begin{tabular}{l|l|l|l}
         &\multicolumn{1}{c}{$E_{tot}$}
         &\multicolumn{1}{c}{$E_c$}
         &\multicolumn{1}{c}{$\% E_c$}  \\
\hline
RPA                         & -0.47207(2)  & -0.20519(56)   &  0.9830(15) \\
\hline
Scaled RPA                  & -0.474825(9) & -0.20794(55)   &  0.9962(15) \\
\hline
Chebyschev                 & -0.474900(9) & -0.20802(55)   &   0.9965(15) \\
\end{tabular}
\end{ruledtabular}
\caption{Total energy $E_{tot}$, correlation energy $E_c$ and percentage of
correlation energy $\% E_c$ for $b=0.1$, $r_s=10$, and $N=22$. The fraction of
the correlation energy recovered is computed from DMC calculations which
provide the exact GS energy for a 1DEG.} 
\label{compenergy}  
\end{table}
We find that the scaled RPA form is essentially equivalent to the true
two-body pseudopotential, obtained with about 10 terms in the sum of Eq.
(\ref{tchebymin}). Moreover, adding a three-body term in $J$, within a
factorization ansatz {\sl \`a la} Feynman retaining the first two $l$-components
does not improve our results in any appreciable manner. 
Therefore in the
following we use a scaled RPA psudopotential optimized for each $N$.

\section{Pair correlations} 
\begin{figure}[t]
\null\vspace{-4.5mm}
\epsfxsize=85mm
\centerline{
\epsffile{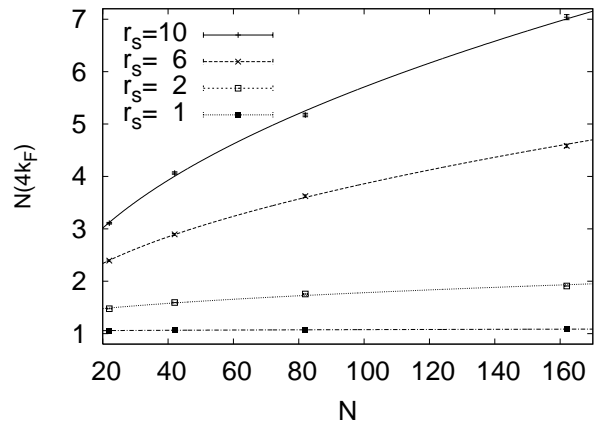}}
\caption{Peak in the charge structure factor  versus the system size. The
  curves are fits to the VMC predictions (symbols).}
\label{peaknum}
\end{figure}

\begin{figure}
\null\vspace{-3mm}
\epsfxsize=85mm
\centerline{
\epsffile{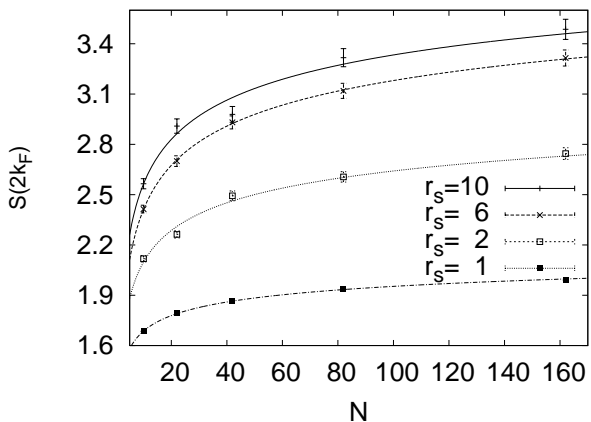}}
\caption{Peak in the spin structure factor  versus the system size.  The
  curves are fits to the VMC predictions (symbols).}   
\label{peakspin}
\end{figure}

According to the predictions of the bosonization technique applied to a
Luttinger Hamiltonian (linearized kinetic energy) with long-range
interaction\cite{schulz}, charge correlations exhibit long range tails
resulting in a divergent peak at $4k_F$ in the charge structure factor
$S_{++}(k)= \langle \rho_+(k) \rho_+(-k) \rangle/N$, whereas spin correlations
decay faster and a finite peak is found at $2k_F$ in the spin structure factor
$S_{--}(k)= \langle \rho_-(k) \rho_-(-k) \rangle/N$. Here, $\rho_{\pm}(k)=
\rho_{\uparrow}(k)\pm\rho_{\downarrow}(k) $ is the Fourier component of the
charge (spin) density.  We have performed VMC simulations for $b=0.1$,
$r_s=1,2,6,10$, $N=10,22,42,82,162$, and computed the above structure factors
at several  $k$ values. In Fig.~\ref{peaknum} and \ref{peakspin} we give the
peak heights $N(4k_F)=S_{++}(4k_F,N)$ and $S(2k_F)=S_{--}(2k_F,N)$, which
appear to be increasing functions of $N$.
As such heights are determined by the long-range behavior of correlations we
would expect them to scale with $N$ similarly to what found for the long-range
Luttinger liquid of \cite{schulz}, for which at large N one readily obtains
\begin{eqnarray}
S(2 k_F) & = & a_3 ( \sqrt{\log{L}} + 1/c ) \exp (- c
\sqrt{\log{L}})+ a_4,
\nonumber \\
N(4 k_F) & = & a_1 L \exp (-4 c \sqrt{\log{L}}) + a_2,
\label{fitpeak}
\end{eqnarray}
with $L=2r_sN$ and $c,a_1$,$a_2$,$a_3$,$a_4$ model and density dependent. We
have used Eqs. \ref{fitpeak} to fit our results, obtaining first $a_1,a_2,c$
from the fit of $N(4 k_F)$ and then $a_3,a_4$ from the fit of $S(2 k_F)$, with
$c$ fixed by the fit of $N(4 k_F)$.  As it can be seen from
Fig.~\ref{peaknum}, \ref{peakspin} and Tab.~\ref{fit}, the long-range tails
implied by Eqs.  \ref{fitpeak} are indeed consistent with our VMC results,
yielding a finite $S(2k_F)$ in the thermodynamic limit. We have reason to
believe that a further improvement of the $\chi^2$ would be obtained by
discarding the smaller values of $N$ in favor of larger ones, on the ground
that Eqs.  \ref{fitpeak} only embody the long-range behavior of correlations.

\begin{table}[t]
\begin{ruledtabular}
\begin{tabular}{l|d|d|d}
         &\multicolumn{1}{c}{$\tilde{\chi}^2_N$}
         &\multicolumn{1}{c}{$\tilde{\chi}^2_S$}
         &\multicolumn{1}{c}{$c$}  \\
\hline
$r_s=1$                 &  4.1  &  3.5  &  0.71 \\
\hline
$r_s=2$                 &  4.7  &  1.5  &  0.78 \\
\hline
$r_s=6$                 &  0.55 &  0.03 &  0.80 \\
\hline
$r_s=10$                &  3.6  &  1.3   & 0.88 \\
\end{tabular}
\end{ruledtabular}
\caption{Fit of the peak of the charge and spin structure factors: reduced
         $\chi^2$  and interaction (fit)  parameter $c$. 
}
\label{fit}  
\end{table}

\section{Summary}
In conclusion, we have shown that a relatively simple (unbiased) optimization
of the wavefunction is able to yield correlations functions for a 1DEG on the
continuum which are in qualitative agreement with the exact predictions of
Schulz \cite{schulz} for a Luttinger Hamiltonian with long-range interactions.
The correlations that we find do support the existence of a quasi Wigner
crystal and a non-ordered unpolarized magnetic phase.  These results are at
variance with what previously found for the same model wire with
VMC\cite{malathesis,malatesta} and DMC\cite{malathesis} simulations, as those
calculations were biased by non-ergodicity.  The results found here are
strongly corroborated by very recent simulations\cite{LRDMC} with a novel
lattice regularized DMC (LRDMC)method\cite{method}, which significantly
alleviates the lack of ergodicity present in the standard DMC.  In particular
we find\cite{LRDMC} that the time projection implemented by the LRDMC only
produces minor quantitative changes in the charge and spin peak heights,
leaving our present conclusions unchanged.

\end{document}